\documentclass[aps,twocolumn,showpacs,amsmath,amssymb,pra,superscriptaddress,floatfix,longbibliography]{revtex4-1}
\usepackage{braket}
\usepackage{amsmath}
\usepackage{amssymb}
\usepackage{mathtools}
\usepackage{graphicx}% Include figure files
\usepackage{dcolumn}% Align table columns on decimal point
\usepackage{bm}% bold math
\usepackage{multirow}
\usepackage{leftidx}
\usepackage{color}
\usepackage{float}
\usepackage{xcolor}
\usepackage{soul}

\usepackage{amsfonts}
\usepackage{eufrak}
\usepackage[german,english]{babel}
\begin{document}
\title{Spatial averaging for light reflection and transmission through
cold atom arrays}
\author{F.~Robicheaux}
\email{robichf@purdue.edu}
\affiliation{Department of Physics and Astronomy, Purdue University, West Lafayette,
Indiana 47907, USA}
\affiliation{Purdue Quantum Science and Engineering Institute, Purdue
University, West Lafayette, Indiana 47907, USA}
\affiliation{ITAMP, Center for Astrophysics $|$ Harvard \& Smithsonian, Cambridge, Massachusetts 02138 USA}

\date{\today}

\begin{abstract}
We theoretically and computationally investigate the role that the spatial
spread of atoms plays in the transmission and reflection of weak light
from atom arrays. In particular, we investigate whether coherent wave
functions for the atoms' positions leads to different results from
a thermal distribution with the same spatial spread. We find that the
coherence is not relevant when the light is weak and the electronic
states evolve on time scales shorter than the oscillation period of
the atoms in their traps. Full numerical calculations and derivations
using the sudden approximation show that reflection and transmission
agree with the simple averaging over atom positions for these
conditions. For parameters
outside these restrictions, the simple spatial averaging may lead to
inaccurate results.
\end{abstract}

\maketitle

\section{Introduction}

Inspired by theoretical and experimental
advances, several groups have investigated 
scenarios involving light interacting with many atoms in regular
arrangements.
\cite{CYL2004,HDC2019,QR12019,CJG2012,JR12012,
SR22016,FJR2016,SWL2017,PBC2017,PBC22017,BMA2017,
AKC2019,NLO2019,ZM12019,MFO2020,JR12019,MPO2019,
WR12020,BLG2020,
BR12020,MAG2020,
BR12021,DC12021,FR12021,BJR2021,
RBY2022,MAG2022,SMA2022,
MHR2022,CMM2022,GAG2022,FFG2022,
WZY2022,PGH2022,WYJ2022,
PZP2023,DPH2023,FFB2023,SS12023,SHM2023,
ORZ2024,WTW2024,BSS2024,
CLO2023,
PWS2021,
AMA2017,MMA2018,GGV2019,BPP2020,CES2020,BOR2022,BR12022,RWP2022,
HCL2023,SPR2024,
MSS2014,GGV2018,BGA2016,CDG2018,WBR2020,RWR2020,GJA2022,SRG2022,ROY2023,
BRS2023,SWH2023,SBS2024,JR12023,SPI2023}
Exciting possibilities arise when the separation of atoms is less than
a wavelength of the light.
Most of the theoretical investigations assume the atoms are on a perfect
grid with zero spatial
deviation.\cite{CYL2004,CJG2012,JR12012,SR22016,FJR2016,SWL2017,AMA2017,PBC2017,
PBC22017,BMA2017,MMA2018,AKC2019,HDC2019,NLO2019,QR12019,ZM12019,MFO2020,JR12019,
GGV2019,MPO2019,WR12020,BLG2020,BR12020,BPP2020,CES2020,MAG2020,BR12021,DC12021,
PWS2021,FR12021,BJR2021,RBY2022,MAG2022,SMA2022,BOR2022,BR12022,RWP2022,MHR2022,
CMM2022,GAG2022,FFG2022,WZY2022,PGH2022,WYJ2022,PZP2023,DPH2023,FFB2023,CLO2023,
SS12023,HCL2023,SHM2023,ORZ2024,WTW2024,BSS2024,SPR2024}
While this is useful for exploring basic concepts, it
is not possible in practice:
the wave function for the center of mass motion must at least
have a spread from the ground state wave function.
For atoms not cooled
to the motional ground state, the spread in positions is larger and
many treatments averaged over spatial
distributions in addition to treating the perfect
lattice.\cite{MSS2014,BGA2016,GGV2018,CDG2018,WBR2020,RWR2020,GJA2022,SRG2022,
ROY2023,JR12023,SPI2023,BRS2023,SWH2023,SBS2024}
There is a wide range of applications for arrays of atoms including
clocks\cite{CYL2004,HDC2019,QR12019},
mirrors, light manipulation, and excitations on a 
lattice\cite{CJG2012,JR12012,BGA2016,SR22016,FJR2016,
SWL2017,PBC2017,PBC22017,BMA2017,CDG2018,AKC2019,NLO2019,ZM12019,MFO2020,JR12019,
MPO2019,WBR2020,WR12020,BLG2020,BR12020,RWR2020,MAG2020,BR12021,DC12021,FR12021,
BJR2021,RBY2022,MAG2022,SMA2022,MHR2022,CMM2022,GAG2022,FFG2022,WZY2022,PGH2022,
GJA2022,SRG2022,WYJ2022,PZP2023,ROY2023,DPH2023,FFB2023,SS12023,BRS2023,SWH2023,
SHM2023,ORZ2024,SBS2024,WTW2024,BSS2024},
collective Lamb shift\cite{MSS2014}, and
quantum information
applications\cite{AMA2017,GGV2018,MMA2018,GGV2019,BPP2020,CES2020,PWS2021,BOR2022,
BR12022,RWP2022,CLO2023,HCL2023,SPR2024}.
Recent overviews cover important aspects of atom arrays.\cite{JR12023,SPI2023}

Some of the calculations treated spatial deviations by fixing the atoms
in space with random positions given by the spatial distribution function
and averaging over the possible positions. While this prescription is
reasonable, there has not been a clear derivation of how the spread
of positions should be treated, especially in the limit the atoms are
cooled to the motional ground state. Does the coherence of the vibrational
wave function require special treatment? At a technical level, are the
results the same for two cases where the spatial part of the density matrix
has diagonal $\rho (x,x)$ the same but off-diagonal
$\rho (x,x')$ substantially different? The purpose of this paper is to
provide a clear derivation and numerical examples of how to treat the
spatial spread of atoms in arrays. The examples we give will be for
when the light and atoms reach steady state although this condition is
not necessary.

For a general treatment,
one of the conditions needs to be that the atoms scatter or
reflect few photons while they evolve into the electronic steady state. If this
condition is not satisfied, the wave function or density matrix associated
with the atoms' positions will evolve. This clearly leads to results that don't
solely depend on the positions of the atoms fixed in space.
Typically, the scattering or
reflection leads to
heating the spatial degrees of freedom although special detuning and
geometry could lead to cooling. This condition can be satisfied
by decreasing the intensity of the incident light. Therefore, except for very
subradiant cases, the approximation will work well when
the one atom Rabi frequency is much less
than the one atom decay rate: $\Omega\ll \Gamma$.

The treatment below will focus on the case where the trap frequency, $\omega_t$,
is much less than the decay rate of the system. Except for very subradiant
cases, this leads to $\omega_t\ll \Gamma$. In typical atom traps, the
trap frequency in the plane of the atom array is different from that
perpendicular to the plane of the atom array. Often, the in-plane frequency
is larger than that perpendicular to the plane, leading to a direction
dependence to $\omega_t$. For the condition $\omega_t\ll \Gamma$, the
largest of the frequencies should be used. 
Most experimental arrangements will satisfy this condition. We
also make this restriction because smaller $\omega_t$ leads to larger
spatial spreads which are more crucial to treat correctly.

We show that within these conditions an accurate,
approximate treatment of the light
plus atom system is to calculate the light properties for atoms fixed in
space and then average the light properties over the positions of the
atoms weighted by their position distribution. We used the sudden
approximation to derive this result, Sec.~\ref{sec:SA}, hence the condition
that the atoms do not move substantially for the duration of the
light-atom interaction. We also
numerically computed the reflection and transmission in simple
cases using a full density matrix treatment, Sec.~\ref{sec:Res}, to
illustrate the accuracy of the fixed in space approximation.
The numerical treatment is the same whether the spread in positions is from
being in a thermal state or being in one vibrational eigenstate.
Restricting the density matrix
calculation to the ground vibrational state is a different
approximation that can lead to qualitatively wrong changes in the
properties of the light (transmission, reflection, etc.).
For example, we describe a case in Sec.~\ref{sec:Res} where even if the recoil
energy is 3,000 times smaller than the vibrational energy spacing,
performing a density matrix calculation with all atoms restricted to
the vibrational
ground state leads to qualitatively wrong changes in the light properties.

\section{Method}

In this section, we describe the method used for the numerical calculation.
In addition, there are
descriptions of the interaction of many atoms with light in a 3D vacuum
and in a 1D waveguide.

\subsection{Full density matrix}

In order to account for the motion of the atoms as well as the
evolution of their internal states, we will use a density matrix
formalism which expands the density matrix in a basis of vibrational
states for the atoms' motion and internal states. We will use the
formalism described in Ref.~\cite{DSR2022} with one exception described
in App.~\ref{sec:dvt}. Also, we will slightly change some of the notation
to avoid confusion of the role of indices.

The calculations will be for $N$ atoms that are trapped in harmonic
wells. To simplify the role of the electronic states, we will only consider
two electronic states for each atom, $j$: $|g_j\rangle$ and $|e_j\rangle$.
These lead to the definition of electronic operators for atom $j$
\begin{equation}
\hat{e}_j\equiv |e_j\rangle\langle e_j| \qquad \hat{\sigma}^-_j\equiv
 |g_j\rangle\langle e_j| \qquad\hat{\sigma}^+_j\equiv
 |e_j\rangle\langle g_j|.
\end{equation}
The position for the center of the atom trap for atom $j$ will be denoted
$\bm{R}_j$ and the operator for the atom position relative to the
trap center will be denoted $\bm{r}_j$. We will assume that the trap
frequency could be different for different directions but we will
denote the angular frequency generically by $\omega_t$. The harmonic
oscillator eigenstate for atom $j$ will be denoted $|n_j\rangle $;
where necessary, the eigenstate for each of the $x,y,z$-directions
will be noted. The light has a wave number $k_0$; when the atoms
interact with a plane light wave the wave vector will be
$\bm{k}_0$. Because the
atoms interact through the electromagnetic field an important position
operator is $k_0\bm{r}_j$ which, for each direction, has the
form $\eta (\hat{a}_j+\hat{a}_j^\dagger )$ with $\hat{a}_j$ the vibrational
lowering operator for the $j$th
atom. The constant $\eta = k_0 \sqrt{\hbar /(2M\omega_t )}$
with $M$ the atom's mass; we are interested in the Lamb-Dicke regime
$\eta\ll 1$ so the spread in positions can be less than the wavelength.

We numerically solve the density matrix equation
\begin{equation}\label{eq:denmat}
\frac{d\hat{\rho}}{dt}=\frac{1}{i\hbar}[\hat{H},\hat{\rho}]+{\cal L}(\hat{\rho})
\end{equation}
with $\hat{\rho}$ the density matrix of the system, $\hat{H}$ is the Hamiltonian,
and ${\cal L}(\hat{\rho})$ is the Lindblad superoperator. The
Hamiltonian consists of three terms
\begin{equation}
\hat{H} = \hat{H}_0 + \hat{H}_L + \hat{H}_{dd}
\end{equation}
where the $\hat{H}_0$ is for the atom traps plus the energy of the electronic
states, the $\hat{H}_L$ is for the laser-atom interaction in the
rotating wave approximation for an incident plane wave with wave vector
$\bm{k}_0$, and $\hat{H}_{dd}$ is for the dipole-dipole
interaction. These terms have the following form
\begin{eqnarray}
\hat{H}_0&=&\hbar\sum_j[\omega_t (\hat{a}_j^\dagger \hat{a}_j + 1/2)
-\Delta \hat{e}_j]\label{eq:H0}\\
\hat{H}_L&=&\hbar\sum_j\frac{\Omega}{2}
\left(e^{i\bm{k}_0\cdot(\bm{R}_j+\bm{r}_j)}\hat{\sigma}_j^+ +
e^{-i\bm{k}_0\cdot(\bm{R}_j+\bm{r}_j)}\hat{\sigma}_j^-
\right)\label{eq:HL}\\
\hat{H}_{dd}&=&\hbar\sum_{j\neq j'}\hat{\Omega}_{jj'}\hat{\sigma}_j^+
\hat{\sigma}_{j'}^-
\end{eqnarray}
where $\omega_t$ is the trap frequency, $\Delta$ is the detuning of the
laser from the transition $\Delta = k_0c - (E_e-E_g)/\hbar$,
$\Omega$ is the Rabi frequency, and $\hat{\Omega}_{jj'}$ is the
position operator dependent part of the dipole-dipole interaction,
Eq.~(\ref{eq:Rabdip}). If the incident light is different from a plane
wave (for example, a focussed beam), then the spatial terms in Eq.~\ref{eq:HL}
are modified. The Lindblad superoperator term is
\begin{equation}\label{eq:lind}
{\cal L}(\hat{\rho})=\sum_{jj'}
\left(
\hat{\sigma}_j^-\hat{\Gamma}_{jj'}(\hat{\rho})\hat{\sigma}_{j'}^+-
\frac{1}{2}\{\hat{\sigma}_{j'}^+\hat{\sigma}_j^-\hat{\Gamma}_{jj'},\hat{\rho}\}
\right)
\end{equation}
where $\{\hat{O}_1,\hat{O}_2\}=\hat{O}_1\hat{O}_2+\hat{O}_2\hat{O}_1$
for any two operators and
$\hat{\Gamma}_{jj'}$ is the position operator dependent part of the
dipole-dipole interaction, Eq.~(\ref{eq:Lindip}).

For atoms interacting in three dimensions far from any surfaces,
the Hamiltonian part of the dipole-dipole operator is
\begin{equation}\label{eq:Rabdip}
\hat{\Omega}_{jj'}=\frac{\Gamma}{2}\left[n_0(s_{jj'})
+\frac{3(\hat{\bm{s}}_{jj'}\cdot\bm{q})(\hat{\bm{s}}_{jj'}
\cdot\bm{q}^*)-1}{2}n_2(\hat{s}_{jj'}) \right]
\end{equation}
where the $n_\ell (x)$ are the Neumann functions,
$\bm{q}$ is the unit vector for the dipole orientation,
$s_{jj'}=k_0|\bm{R}_j+\bm{r}_j-\bm{R}_{j'}-\bm{r}_{j'}|$
and $\hat{s}_{jj'}$ is the unit vector for the atom difference.
Because of the presence of the $\bm{r}_j$ and $\bm{r}_{j'}$,
the $\hat{\Omega}_{jj'}$ is an operator that can change the
vibrational quantum numbers of atoms $j$ and $j'$.
The Lindbladian part of the dipole-dipole operator is
\begin{equation}\label{eq:Lindip}
\hat{\Gamma}_{jj'}=\Gamma\left[j_0(s_{jj'})
+\frac{3(\hat{\bm{s}}_{jj'}\cdot\bm{q})(\hat{\bm{s}}_{jj'}
\cdot\bm{q}^*)-1}{2}j_2(\hat{s}_{jj'}) \right]
\end{equation}
with the $j_\ell (x)$ the spherical Bessel functions.
Using
$h_0^{(1)}(s)=e^{is}/[is]$ and $h_2^{(1)}(s) = (-3i/s^3 - 3/s^2 + i/s)e^{is}$
$j_\ell (s)=\Re [h^{(1)}_\ell (s)]$ and
$n_\ell (s)=\Im [h^{(1)}_\ell (s)]$. Similar to the $\hat{\Omega}_{jj'}$,
the $\hat{\Gamma}_{jj'}$ is an operator due to the presence of the
$\bm{r}_j$ and $\bm{r}_{j'}$ and can change vibrational quantum numbers.
The trickiest part is the first term of Eq.~(\ref{eq:lind}) where the
$\bm{r}_j$ acts from the left on the density matrix while the
$\bm{r}_{j'}$ acts from the right on the density matrix.
See the more detailed discussion in Refs.~\cite{FRH2019,DSR2021,DSR2022}.

For atoms that can only emit light into or absorb from a one dimensional
wave guide, we will take the Hamiltonian part of the dipole-dipole operator
as
\begin{equation}\label{eq:O1D}
\hat{\Omega}_{jj'}=\frac{\Gamma}{2}\sin(k_0|X_j+x_j-X_{j'}-x_{j'}|)
\end{equation}
and
\begin{equation}\label{eq:G1D}
\hat{\Gamma}_{jj'}=\Gamma\cos (k_0 [X_j+x_j-X_{j'}-x_{j'}]).
\end{equation}
While these equations are simpler than those for light in three
dimensions, they lead to the same basic physical processes:
transmission and reflection of light, vibrational excitation or
de-excitation, etc.

We solve for the time dependent density matrix using the eigenstates of
the $\hat{H}_0$ operator, Eq.~(\ref{eq:H0}). We will define the state
$|\alpha\rangle$ to be the tensor product of individual atom eigenstates.
For atom $j$, we use $i_j$ to be 0 or 1 for $|g_j\rangle$ or $|e_j\rangle$
and $|n_j\rangle$ the vibrational state of atom $j$ which will be limited to
the range 0 to $n_{max}=N_{vib}-1$. This gives
\begin{equation}
|\alpha\rangle \equiv |i_1\rangle\otimes |n_1\rangle\otimes |i_2\rangle\otimes
|n_2\rangle \otimes ...
\end{equation}
where there are $(2N_{vib})^N$ states altogether.
We define the density matrix as
\begin{equation}\label{eq:vrep}
\hat{\rho}=\sum_{\alpha\alpha '} |\alpha\rangle \rho_{\alpha\alpha '}
\langle\alpha '|.
\end{equation}
To evaluate an operator $\hat{O}$ acting on $\rho$, we use the representation
\begin{equation}
\langle\alpha |\hat{O}\hat{\rho}|\alpha '\rangle=\sum_{\alpha ''}
\langle\alpha |\hat{O}|\alpha ''\rangle\langle\alpha ''|\hat{\rho}|\alpha '\rangle
\end{equation}
which is relatively efficient because the operators in Eq.~(\ref{eq:denmat})
are extremely sparse.

\subsection{Transmitted and reflected light}\label{sec:tandr}

Since we are interested in the effect of atoms' positions on the interaction
of light, we also need to have a form for the operators that can be
evaluated to compute the intensity of light at a position away
from the atoms. The transmitted and reflected light in steady state is
somewhat undefined because for most cases the interaction with the light
will lead to the atoms' motion heating with time. Strictly speaking, there
is no steady state. However, in most experiments, it is assumed that the
light is weak enough that the internal states of the atom reach steady
state before vibrations appreciable change. Thus, we will investigate
the reflection and transmission on this time scale by choosing the Rabi
frequency to be much smaller than $\Gamma$ in most of our examples.

For light in three dimensions,
the calculation of the electromagnetic flux in the direction
$\bm{\mu} $ at position $\bm{R}$ involves an electric field
type operator with a form
\begin{equation}
\hat{\cal\bm E}(\bm{R})={\cal\bm E}_{cl}(\bm{R}) + \sum_j\hat{\sigma}^-_j
\bm{g}_E(\bm{R}-\bm{R_j}-\bm{r_j})
\end{equation}
where $\bm{g}_E(\bm{R})\propto [\bm{q}-\hat{\bm{R}}(\hat{\bm{R}}\cdot \bm{q})]
 h^{(1)}(kR)$ in the far field
and a magnetic field type operator with a form
\begin{equation}
\hat{\cal\bm B}(\bm{R})={\cal\bm B}_{cl}(\bm{R}) + \sum_j\hat{\sigma}^-_j
\bm{g}_B(\bm{R}-\bm{R_j}-\bm{r_j})
\end{equation}
where $\bm{g}_B(\bm{R})=\hat{\bm{R}}\times \bm{g}_E(\bm{R})$ in the far
field. The ${\cal\bm E}_{cl}(\bm{R})$ and ${\cal\bm B}_{cl}(\bm{R})$
are proportional to the classical electric and magnetic fields from the
laser. The flux in the direction $\bm{\mu}$ then has the form
\begin{equation}
{\cal F}\propto \mu\cdot Tr[\hat{\cal\bm E}^\dagger
(\bm{R})\times \hat{\cal\bm B}(\bm{R})
\hat{\rho}-\hat{\cal\bm B}^\dagger
(\bm{R})\times \hat{\cal\bm E}(\bm{R})
\hat{\rho}]
\end{equation}
There are simpler equations for the far field involving the sum over the
dipoles with appropriate phase factors. Because there are position operators
in this expression, there are density matrix terms off diagonal in
vibrational quantum numbers which contribute. The transmission and
reflection probabilities can be calculated from the flux.

For light in one dimension, the equations simplify so that the
operator
\begin{equation}
\hat{\tau}(x_1,x_2,...)
=1-i\frac{\Gamma}{\Omega}\sum_j\hat{\sigma}^-_je^{-ik_0(X_j+x_j)}
\end{equation}
leads to the transmission probability
\begin{equation}\label{eq:tran}
T = Tr[\hat{\tau}^\dagger \hat{\tau}\hat{\rho}]
= Tr[\hat{\tau}\hat{\rho}\hat{\tau}^\dagger ]
\end{equation}
and the operator
\begin{equation}
\hat{\theta}(x_1,x_2,...)
=-i\frac{\Gamma}{\Omega}\sum_j\hat{\sigma}^-_je^{ik_0(X_j+x_j)}
\end{equation}
leads to the reflection probability
\begin{equation}\label{eq:refl}
R = Tr[\hat{\theta}^\dagger \hat{\theta}\hat{\rho}]
= Tr[\hat{\theta}\hat{\rho}\hat{\theta}^\dagger ].
\end{equation}
Note that there are the position operators, $x_j$, in both the transmission
and reflection probability which can lead to connections with off-diagonal
vibrational states in the density matrix.

\section{Sudden Approximation}\label{sec:SA}

The sudden approximation can be quite accurate
when the electronic states evolve on time scales much faster than the
oscillation period of the atoms. Also, the atoms need to scatter or
reflect few enough photons that they are not appreciably accelerated
from the recoil.
The sudden approximation was used in Refs.~\cite{FRH2019,DSR2021} and
verified in Ref.~\cite{DSR2022} for photons interacting collectively
with many atoms.

To explore the sudden approximation, we first transform the density
matrix from the vibrational basis into the position basis using the
vibrational wave functions:
\begin{eqnarray}
&\null &
\langle \bm{r}_1,\bm{r}_2,...|\hat{\rho}|\bm{r}'_1,\bm{r}'_2...\rangle
=\sum_{n_1,n_2...}\sum_{n'_1,n'_2...}\psi_{n_1}(\bm{r}_1)\psi_{n_2}(\bm{r}_2)
...\nonumber\\
&\null &\langle n_1|\langle n_2|...\langle n_N|
\hat{\rho}|n_1'\rangle |n_2'\rangle ...|n'_N\rangle
\psi_{n'_1}(\bm{r}'_1)\psi_{n'_2}(\bm{r}'_2)
...
\end{eqnarray}
with $\psi_{n_j}(\bm{r}_j)$ the vibrational wave function for atom $j$.
At this point, there are no approximations. This form is much more difficult
to use in a fully numerical treatment because many more points in real
space, $\bm{r}_1,\bm{r}_2...$, are required for converged calculations
compared to the vibrational states of Eq.~(\ref{eq:vrep}).
The trapping Hamiltonian in Eq.~(\ref{eq:H0})
leads to coherences between $\bm{r}_j$ and $\bm{r}'_j$ and the transfer
of amplitude between different points in the $3N$ space. To be clear,
from completeness
arguments, this formulation must give the same result when
converged as using the vibrational
basis functions, Eq.~(\ref{eq:vrep}).

The basic idea for the sudden {\it approximation}
is to disregard the trapping potential
and atom kinetic energy in the Hamiltonian.
This still leads to a
density matrix with off diagonal coherence for the positions ${\bf r}_j$
and ${\bf r}'_j$ in the left and right sides. Also, the amplitudes at
different positions evolve differently due to the position
dependence of the photon-atom and dipole-dipole interactions.
However, it does not involve amplitudes moving to different positions
and, thus, calculations for separate values of
$\bm{r}_1,\bm{r}_2,...,\bm{r}'_1,\bm{r}'_2...$ can be performed
independently.
This method can be used to approximately calculate the recoil delivered to the
atoms, giving accurate values when $\omega_t\ll\Gamma$. In fact,
it is precisely these types of terms investigated in Refs.~\cite{FRH2019,DSR2021}
that lead to changes in the average kinetic energy of the atoms which
was verified in Ref.~\cite{DSR2022}.

The position representation of the density matrix leads to simple forms
of the flux, transmission probability, and reflection probability. To
see this, examine the expression for the transmission probability,
Eq.~(\ref{eq:tran}), in the position representation:
\begin{eqnarray}\label{eq:trpos}
T &= &\int dx_1dx_2...\hat{\tau}^\dagger (x_1,x_2,...)
\hat{\tau} (x_1,x_2,...)\times\nonumber\\
&\null &\langle x_1,x_2...|\hat{\rho}|x_1,x_2,...\rangle .
\end{eqnarray}
There are no approximations in this expression.

Suppose the system starts in the ground electronic state with any type
of positions coherence in the density matrix
\begin{equation}
\langle x_1,x_2...|\hat{\rho}|x_1',x_2',...\rangle =
\rho_0 (x_1,x_2,...,x_1',x_2'...)|{\cal G}\rangle\langle {\cal G}|
\end{equation}
with $|{\cal G}\rangle$ is all electronic states in the ground state
and the $\rho_0$ function encapsulating any positional coherences.
Within the sudden approximation, the $\rho_0$ does not change and
the electronic part of the density matrix evolves depending on the positions.
Once the system reaches steady state, the electronic part goes to
\begin{equation}
|{\cal G}\rangle\langle {\cal G}| \to \hat{\rho}_{SA}(x_1,x_2,...,x'_1,x'_2,...)
\end{equation}
with the sudden approximation density matrix the same as that calculated in
Refs.~\cite{FRH2019,DSR2021}. Using this in the expressions for the
flux, transmission probability, or reflection probability leads to a
conceptually simple reduction. For example, the transmission probability
can be written as
\begin{eqnarray}\label{eq:trSA}
T&= &\int dx_1dx_2...\hat{\tau}^\dagger (x_1,x_2,...)
\hat{\tau} (x_1,x_2,...)\times\nonumber\\
&\null &\rho_0(x_1,x_2,...,x_1,x_2,...)
\hat{\rho}_{SA}(x_1,x_2,...,x_1,x_2,...)\nonumber\\
&=&\int dx_1dx_2... P(x_1,x_2,...) T_{SA}(x_1,x_2...)
\end{eqnarray}
where $P(x_1,x_2,...)=\rho_0(x_1,x_2,...,x_1,x_2,...)$ is the initial
probability density for finding atom 1 at position $x_1$ and
atom 2 atom position $x_2$ ... and the sudden approximation
transmission probability
\begin{eqnarray}
T_{SA}(x_1,x_2...) &= &\hat{\tau}^\dagger (x_1,x_2,...)
\hat{\tau} (x_1,x_2,...)\times\nonumber\\
&\null &
\hat{\rho}_{SA}(x_1,x_2,...,x_1,x_2,...)
\end{eqnarray}
is the probability for photon transmission if atom 1 is at position
$x_1$ and atom 2 is at position $x_2$ etc. A similar treatment of
the flux and reflection probability leads to the same form: compute
the relevant quantity (flux, transmission probability, reflection
probability, ...) for atoms fixed in space and average over their
possible positions from $P(x_1,x_2,...)$.

Although this result is simple, it does lead to somewhat
surprising conclusions. For example, suppose all atoms start in the vibrational
ground state and the parameter $\eta = 
k_0\sqrt{\hbar/(2M\omega_t)}$ is tiny. Since the recoil energy over
the vibrational spacing is $\eta^2$, one might expect that
one could use the approximation that the atoms are in the ground
state throughout the evolution. As will be seen in the Results section,
this can qualitatively miss the changes due to the spread in positions.

\subsection{Evolution time scale}

The discussion of the sudden approximation depends on the duration of the
light-atom interaction. Sometimes the situation involves the properties
of the light once the atoms reach steady state in which case the duration
is until the electronic states stop evolving. Sometimes the properties
are required before steady state is reached which would shorten the
duration. The accuracy of the sudden
approximation requires the atoms to be fixed in space for this duration.
By this we mean, the atom positions would not be
expected to change in an important way on the time scale of the
measurement of the light properties.

Weak light, $\Omega\ll\Gamma$,
leads to the simplest case. For steady state properties,
the electronic states will require several lifetimes. The relevant lifetime for
light coupling to superradiant states is less than $1/\Gamma$ while
light coupling to subradiant states leads to relevant lifetimes larger
than $1/\Gamma$. For this case, the condition will often be
$(10-100)/\Gamma \ll 1/\omega_t$ depending on the extent of subradiance.

If the light is intense and/or the $\eta$ becomes large, the recoil of the
atoms or dipole-dipole forces
can also become important. The recoil or the dipole-dipole interactions
could cause the atoms to accelerate so that the atom
velocity increases substantially.
The atom separation divided by the increased speed then becomes
the comparison time scale, not $1/\omega_t$.

Lastly, we restrict the atoms to be cold. If, the atoms could be initially
moving quickly, i.e. higher temperature, then
the electronic states need to evolve on time scales fast compared
the atom separation divided by the speed of the atoms. For this
case, Doppler effects could lead to the collective light-atom interaction being
relatively uninteresting.

\section{Results}\label{sec:Res}

While the derivation of the previous section clearly shows that for the
sudden approximation
case, the proper procedure is to compute the flux, reflection probability,
or transmission probability by fixing the atoms in space
and then averaging over their positions, it is interesting to numerically
examine a case that can show this result. In particular, we want to show that
using the approximation that all atoms are in the ground vibrational
state throughout the calculation is a qualitatively different, and inaccurate,
approximation.

For atoms in free space, $N\gg 1$
to clearly distinguish between large scale transmission
and reflection. This leads to very large calculations because there will
need to be at least a few vibrations for every atom. For $N$ atoms with
two internal states and $n_{max}=N_{vib}-1$ the maximum vibrational quantum,
there are $[2N_{vib}]^{2N}$ terms in the density matrix.
Fortunately, the one-dimensional example can have transmission and
reflection probabilities that strongly depend on the atom separations
for two atoms
when the average separation is near an integer number of wavelengths.
This allows a fully numerical simulation with modest computational resources.

The one dimensional case with two atoms near an integer wavelength separation
has a subradiant interference feature leading to a relatively sharp transmission
peak. The position and width of the transmission peak depend strongly on the actual
separation so a spread in atom positions can lead to a clear change.

\begin{figure}
\resizebox{80mm}{!}{\includegraphics{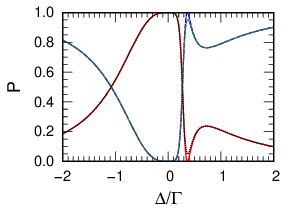}}
\caption{\label{fig:avg}
The transmission, $T$, and reflection, $R$, probabilities for the one
dimensional case discussed in the text. For the case where the separation, $d$,
of the atoms is exactly $0.9\lambda$, the red solid (blue long dash) line
is the $R$ ($T$) probability. When the average separation, $\langle d\rangle$,
is $0.9\lambda$
but the positions are averaged over the vibrational ground state, the
black dotted (green dash) is the $R$ ($T$) probability.
}
\end{figure}

\begin{figure}
\resizebox{80mm}{!}{\includegraphics{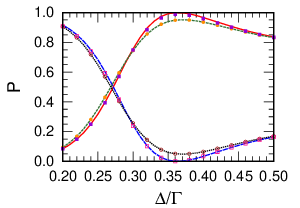}}
\caption{\label{fig:vib}
Same as Fig.~\ref{fig:avg} but also showing the results of the restricted
quantum calculation with $n_{max}=0$
(purple solid square for $T$, empty pink square for $R$)
and the converged quantum calculation restricted with $n_{max}=4$ for
both atoms (orange solid circle for $T$, maroon empty circle
for $R$).
}
\end{figure}

The results in Figs.~\ref{fig:avg} and \ref{fig:vib} are for two atoms
interacting through a one-dimensional wave guide,
Eqs.~(\ref{eq:O1D}) and
(\ref{eq:G1D}). To ensure that we satisfied the conditions for the
sudden approximation, we chose somewhat extreme parameters:
$\Gamma = 2\pi 10$~MHz, $\Omega=10^{-4}
\Gamma $, $\lambda=3.1$~mm, $d=0.9\lambda$, $M=1.6605\times 10^{-28}$~kg,
$\omega_{t}=10^3$~s$^{-1}$. This leads to $\eta\simeq 0.036$ and a recoil
energy over vibrational spacing of $\eta^2\simeq 0.0013$.
Because there are two atoms in their ground vibrational state, the
standard deviation of the separation is $\sigma = \sqrt{\hbar /(M\omega_t)}$
leading to $\sigma /\lambda = \eta /(\pi\sqrt{2})\simeq 0.0081$.
Figure~\ref{fig:avg} compares the transmission, $T$, and reflection, $R$,
probabilities versus
detuning for the perfect separation and from averaging
over the positions of the atoms in their vibrational ground state.
For most of the range of detuning, $\Delta$, averaging over the atoms'
positions has little effect. The largest effect from averaging
is near $\Delta = 0.36\Gamma$
where the interference leads to a sharp peak in the transmission probability.
The effect is large for that detuning
because the position and width of the transmission peak strongly depends
on the value of the separation when the separation $d\sim \lambda$.

Figure~\ref{fig:vib} is the same plot over a smaller range of detuning
with the results from two full quantum calculations,
Eqs.~(\ref{eq:tran}) and (\ref{eq:refl}), with different
number of vibrational states. The quantum calculation that restricted
all vibrations to be in the ground state, $n_{max}=0$, has the
purple solid square for $T$ and the empty pink square for $R$.
These values
are in decent agreement with the calculation that {\it does not average}
over the atoms' positions. However, the transmission plus reflection
probabilities for this approximation only adds up to 0.990 at
$\Delta = 0.36\Gamma$ indicating there is a lack of convergence. By
increasing $n_{max}$, we can demonstrate convergence of the quantum
calculation. In Fig.~\ref{fig:vib}, the $n_{max}=4$ results are
orange solid circle for $T$ and maroon empty circle for $R$
which are in excellent agreement with the simple spatial
averaging. The largest difference is near $\Delta =0.36\Gamma$.
At this value of detuning, the sum of reflection and transmission probabilities
give $1-R-T\simeq 4.6\times 10^{-7}$, and
the difference between the simple spatial averaging and
the quantum calculation of the transmission probability
was $\sim 3\times 10^{-6}$. This demonstrates that even though the ratio of
recoil energy to vibrational energy spacing is tiny, $\eta^2\sim 0.0013$,
one can not restrict the calculation to the vibrational ground state.

We performed calculations for $d=0.95\lambda$ and
$\omega_t = 4\times 10^3$~s$^{-1}$ with everything else kept the
same. This decreases the change from
integer lambda spacing by a factor of 2 from 0.1 to 0.05 which
makes the subradiant state lifetime approximately 4 times longer.
By increasing $\omega_t$ by a factor of 4, the $\eta$ decreases by
a factor of 2. This keeps the $\sigma /\Delta\lambda$ the same.
We found the same trends as in Fig.~\ref{fig:vib}
with the $n_{max}=0$ in good agreement with the $d=0.95\lambda$
curve and the $n_{max}=4$ in excellent agreement with the spatially
averaged results. This case has a larger change in probabilities
due to averaging:
the maximum for $T$ after averaging is $\simeq 0.842$. The $n_{max}=0$
calculation has a maximum for $T$ of 0.98.
Note that decreasing $\eta$ means the ratio of recoil
energy to vibrational energy spacing decreased by a factor of 4 to
$\eta^2\simeq 3.3\times 10^{-4}$ which might make it even more
surprising that the $n_{max}=0$ calculation gets the change in
probabilities so wrong.

\begin{figure}
\resizebox{80mm}{!}{\includegraphics{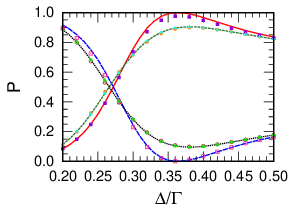}}
\caption{\label{fig:vib2mm}
Similar to Fig.~\ref{fig:vib} but for $\lambda = 2$~mm which increases
the relative size of the averaging region. The red solid (blue long dash)
line is the  $R$ ($T$) probability for perfect separation while the
black dotted (green dash) spatially average the positions.
The results of the $n_{max}=0$ (purple solid square for $T$, empty pink square
for $R$), $n_{max}=1$ (orange solid circle for $T$, maroon empty circle
for $R$),
and $n_{max}=2$ (turquoise solid triangle for $T$, empty green triangle for $R$).
}
\end{figure}

Changing the wavelength to $\lambda = 2$~mm, leads to $\eta \simeq 0.056$
and $\sigma/\lambda \simeq 0.0126$. This gives a spread of positions that
is 1.55~times larger than the calculations in Figs.~\ref{fig:avg} and
\ref{fig:vib} which will lead to a larger effect from averaging.
The results are shown in Fig.~\ref{fig:vib2mm} where it is clear that
spatial averaging has a larger effect. In the plot are shown
the points for calculations with $n_{max}=0$, 1, and 2 where it is clear that
most of the change occurs when going from 0 to 1. In fact for both the
$\lambda = 3.1$ and 2~mm calculations, the $n_{max}=1$ results are good enough
to reach better than 1\% accuracy. For the 2~mm calculation at
$\Delta = 0.36\Gamma$, going from
$n_{max}=0$ to 1 to 2 gives $1-R-T$ going from $2.4\times 10^{-2}$
to  $2.3\times 10^{-3}$ to $3.7\times 10^{-4}$ and an error in $T$ going
from $8.0\times 10^{-2}$ to  $-6.4\times 10^{-3}$ to $5.4\times 10^{-4}$.

\begin{figure}
\resizebox{80mm}{!}{\includegraphics{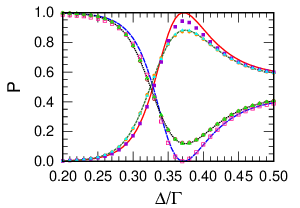}}
\caption{\label{fig:vib32mm}
Same as Fig.~\ref{fig:vib2mm} but for 3 atoms with a separation
of $d=0.85\lambda$.
}
\end{figure}

We repeated the calculations that used $\lambda = 2$~mm
but with 3 atoms to ensure there was nothing
special about the 2 atom case. Because there are more atoms, the separation
does not need to be as close to an integer wavelength to get sensitivity
to the atom positions. In Fig.~\ref{fig:vib32mm}, we show the transmission
and reflection probability versus detuning for 3 atoms with average
separation of $d=0.85\lambda$. The line and point types are the same
as Fig.~\ref{fig:vib2mm}. As with the previous plot, these
show the very strong error for the calculation with $n_{max}=0$ and
progressive convergence toward the spatial averaging result with
increasing $n_{max}$.
At $\Delta = 0.36\Gamma$, going from
$n_{max}=0$ to 1 to 2 to 3 gives $1-R-T$ going from $6.3\times 10^{-2}$
to  $3.1\times 10^{-3}$ to $7.7\times 10^{-4}$ to $1.3\times 10^{-4}$
and an error in $T$ going
from $5.4\times 10^{-2}$ to  $-5.3\times 10^{-3}$ to  $2.0\times 10^{-4}$
to  $-6.6\times 10^{-5}$.

\begin{figure}
\resizebox{80mm}{!}{\includegraphics{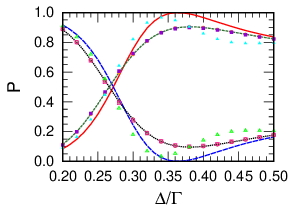}}
\caption{\label{fig:vib2mmwt}
Similar to Fig.~\ref{fig:vib2mm} but for different $\omega_t$ with the
$M$ inversely changed to keep $\sigma = \sqrt{\hbar /(M\omega_t)}$ unchanged.
The results of the $n_{max}=4$ for
$\omega_t=10^5$~s$^{-1}$ (orange solid circle for $T$, maroon empty circle
for $R$), $10^6$~s$^{-1}$ (purple solid square for $T$, empty pink square for $R$),
and $10^7$~s$^{-1}$ (turquoise solid triangle for $T$, empty green triangle
for $R$).
}
\end{figure}

The sudden approximation can be made to fail by having the trap frequency,
$\omega_t$, increase. However, changing the trap frequency alone causes
other basic parameters to change. Importantly, increasing $\omega_t$
decreases the spread of the wave function. One could keep the spread
fixed by
inversely changing the mass, $M$. For example, increasing $\omega_t$ by
a factor of 10 and decreasing $M$ by a factor of 10.
Another method would be to change
$\Gamma$ and $\Omega$ keeping their ratio fixed. Both methods lead to
the same reflection and transmission probabilities when $\omega_t/\Gamma$
is the same in each method.
Figure~\ref{fig:vib2mmwt}
shows the result for $\lambda = 2$~mm and $\eta \simeq 0.056$.
The $\omega_t = 10^5$ and $10^6$~s$^{-1}$ give nearly
the same result as the previous case where $\omega_t = 10^3$~s$^{-1}$
which closely matches the spatial average. However, the
$\omega_t = 10^7$~s$^{-1}$ result is substantially different.
For this trap frequency, the decay time, $1/\Gamma$ is still a small
fraction of the trap period, $2\pi /\omega_t$. However, the proper
comparison is between the trap period and the lifetime of the subradiant
excitation which are comparable.

The sudden approximation also can be made to fail for larger values of the
Rabi frequency, $\Omega$. We repeated the calculations for the parameters
of Fig.~\ref{fig:vib2mm} but with $\Omega = 10^{-3}$, $0.01$, and $0.1\Gamma$.
All of these cases have $\Omega\gg \omega_t$, and, for the fixed atom calculations, lead to steady state results for times much less than $1/\omega_t$.
Except for the
$\Omega =\Gamma /10$ case, the simple spatial averaging gave an
excellent approximation to the converged vibrational calculation.
However,
the $\Omega =\Gamma /10$ case gave qualitatively different results between
the simple spatial averaging and the converged vibrational calculation
because the atoms substantially shift positions under the radiation
pressure from the reflected and scattered photons.

\section{Summary}

We have explored the role that spatial averaging plays in the reflection
or transmission of photons through atom arrays. In experiments, the
atoms will be held in space using trapping lasers leading to a spread
in atom positions which affects the transmission and reflection properties.
We envision situations where the light intensity is
low enough that the trap energy of the atoms hardly changes.

Using the sudden approximation and fully numerical calculations, we have
shown that simply averaging the reflection and transmission over the atom
positions is accurate when the trap period is much larger than the evolution
time scale of the electronic states. This approximation breaks down
when the trap period is comparable to or shorter than electronic state
evolution time scale or for high intensity where the atom velocity
and position change due to radiation pressure or dipole-dipole forces.

Data plotted in the figures is available at~\cite{data}.

\begin{acknowledgments}
I am grateful to Chen-Lung Hung for suggesting this question and discussions
with Deepak Suresh.
This work was supported by the U.S. National Science Foundation under Grant
No. 2410890-PHY and
through a
grant for ITAMP at Harvard University.
\end{acknowledgments}

\appendix

\section{Evaluation of vibrational matrix elements}\label{sec:dvt}

The differential equations for the density matrix requires the evaluation
of matrix elements of complicated functions involving different vibrational
levels. In the limit that $\eta$ is small, these matrix elements are often
evaluated using a power series technique. For example, evaluating a matrix
element of a function of the operator $\hat{x}$ could use an approximation like
\begin{equation}
\langle n |F(X + \hat{x})|n'\rangle =F(X) \langle n|n'\rangle
+ F'(X) \langle n|\hat{x}|n' \rangle
\end{equation}
which stops at first order in the expansion. A better approximation would
continue to second (or higher) order in $\hat{x}$. However, when $F$ is a
complicated function, the evaluation of higher derivatives can be complicated
and are often discarded for third and higher order.

We use a discrete variable method for evaluation of matrix elements of complicated
functions.\cite{LPL1982} The idea is to find the eigenvalues and eigenvectors
of the matrix of the $\hat{x}$ operator and use those to evaluate the matrix
elements. For a finite basis set from 0 to $n_{fin}$, find
\begin{equation}
\sum_{n'=0}^{n_{fin}}
\langle n|\hat{x}|n' \rangle U_{n'\beta} = U_{n\beta}x_\beta
\end{equation}
where $U$ is the unitary matrix of eigenvectors and $x_\beta$ are the
eigenvalues. The matrix elements are approximately given by
\begin{equation}
\langle n |F(X + \hat{x})|n'\rangle\simeq \sum_\beta U_{n\beta}
F(X+x_\beta )U^\dagger_{\beta n'}
\end{equation}
where the accuracy increases for larger $n_{fin}$.

There are many advantages of this method of which we will give three:
1) the computation of the derivatives
of $F$ is unnecessary. 2) if $F$ is a unitary operator,
the resulting matrix will be exactly unitary if $n_{fin}$ matches the
$n_{max}$ in the calculation. 3) for harmonic oscillator basis functions
and the $\hat{x}$ operator, the $\langle n|\hat{x}|n' \rangle$ is tridiagonal
leading to the equivalent of high order Taylor series expansion of $F$ for
the small $n,n'$ matrix elements.

\bibliography{ArrayVibrRT}

\end{document}